\documentclass[twocolumn,amsmath,amssymb,nofootinbib,tighten,prl]{revtex4-1} 

\usepackage{graphicx}
\usepackage{dcolumn} 
\usepackage{bm, bbm}      
\usepackage{curves}
\usepackage{epic}
\usepackage{wasysym}
\usepackage{epsf, times}
\usepackage{subfigure}
\usepackage{amsthm}

\def\openone{\leavevmode\hbox{\small1\kern-4.2pt\normalsize1}}
\def\DTO{Dy$_2$Ti$_2$O$_7$}
\def\HTO{Ho$_2$Ti$_2$O$_7$}

\def\K{\mathcal{K}}

\def\P{\mathcal{P}}

\def\unue#1{{\it#1 --}}

\newcommand{\beq}{\begin{equation}}
\newcommand{\eeq}{\end{equation}}
\newcommand{\bea}{\begin{eqnarray}}
\newcommand{\eea}{\end{eqnarray}}
\newcommand{\bfig}{\begin{figure}}
\newcommand{\efig}{\end{figure}}
\newcommand{\bei}{\begin{itemize}}
\newcommand{\eei}{\end{itemize}}

\begin{document}


\title{
Thermal quenches in spin ice
}

\author{
C. Castelnovo$^1$ 
}
\author{
R. Moessner$^2$
}
\author{
S. L. Sondhi$^3$
}
\affiliation{
$^1$ 
Rudolf Peierls Centre for Theoretical Physics and Worcester College, 
Oxford University, Oxford OX1 3NP, UK
}
\affiliation{
$^2$ 
Max-Planck-Institut f\"ur Physik komplexer Systeme, 
N\"othnitzer Stra{\ss}e 38, 01187 Dresden, Germany 
}
\affiliation{
$^3$ 
Department of Physics, Princeton University, 
Princeton, NJ 08544, USA 
}

\date{\today}

\begin{abstract}
We study the diffusion annihilation process which occurs when spin ice
is quenched from a high temperature paramagnetic phase deep into
the spin ice regime, where the excitations -- magnetic 
monopoles -- are sparse. We find that due to the Coulomb interaction
between the monopoles, a dynamical arrest occurs, in which non-universal 
lattice-scale constraints impede the complete decay of charge fluctuations. 
This phenomenon is outside the reach of conventional mean-field theory 
for a two-component Coulomb liquid. 
We identify the relevant timescales for the dynamical arrest and propose an
experiment for detecting monopoles and their dynamics in spin ice
based on this non-equilibrium phenomenon. 
\end{abstract}

\maketitle
%
%

\unue{Introduction} 
There is intense current interest in the study of strongly correlated systems 
hosting fractionalised excitations, in fields as diverse as magnetism, quantum 
Hall physics and quantum computing, or even the study of (topological) band 
insulators. Such excitations arise against the background of highly unusual
ground states. 

Recently, we have argued that spin ice -- an Ising magnet on the pyrochlore 
lattice -- hosts deconfined magnetic monopole excitations, which result from 
the fractionalisation of the high-energy local dipole
moments~\cite{Castelnovo2008}. At the moment, the focus of theoretical and
experimental studies consists of predicting and detecting signatures of these 
excitations~\cite{Jaubert2009,Morris2009,Fennell2009,Kadowaki2009,Bramwell2009}. 

In spin ice, the ground state ensemble is unusual in that it exhibits algebraic 
correlations without representing a conventional critical point: this Coulomb 
phase -- in the sense of the deconfined phase of a $U(1)$ gauge theory -- is a 
consequence of the local constraint that two spins point into each 
tetrahedron and two point out. Indeed, spin ice owes its name to this magnetic 
version of the Bernal-Fowler ice rules. The Coulomb phase is characterised by 
an emergent gauge field, rather than an emergent order parameter; 
as such, it is a classical example of topological order. 

Violating the ice rules by flipping a spin out of a ground-state configuration,
at a cost in energy of $\Delta_{\rm sf}$, leads to {\em a pair} of pointlike
defects in the tetrahedra the spin belongs to. These two defects are
deconfined: they can be separated to an arbitrarily large distance at a finite
cost in energy. In the presence of long-range dipolar interactions -- the model
referred to as dipolar spin ice below -- such defects experience a magnetic
Coulomb interaction, $V(r)=\mu_0 Q_m^2/(4 \pi r)$, whence the appellation 
\emph{magnetic monopoles}. 
Here, $\mu_0$ is the vacuum permeability, and the magnetic charge 
$Q_m=2 |\vec{\mu}|/a_d$ is related to the dipole moment of the magnetic ions, 
$|\vec{\mu}|$, and the distance between the centres of adjacent tetrahedra, 
$a_d$. 

In addition, there is a Coulomb interaction of entropic origin, with 
coupling strength $Q_s^2 \propto T$. It is present 
even in the nearest-neighbour model for spin ice, 
where the long-range dipolar interactions are omitted and $Q_m=0$. 

Spin ice compounds, such as {\HTO} and {\DTO}, are thus the first 
instances of three-dimensional magnets which host deconfined fractionalised 
excitations. The most satisfying detection experiment would consist of a 
direct visualisation of a magnetic monopole in bulk spin ice. 
However, due to its small magnetic charge and the fact that single 
quasiparticles are hard to come by in bulk systems -- even in quantum Hall 
physics, a single fractionalised charge has never been imaged -- 
this has so far proven beyond reach. 
In Ref.~\onlinecite{Castelnovo2008}, we have 
shown that a thermodynamic signature of the magnetic Coulomb interaction of 
the monopoles is the presence of a liquid-gas transition in a 
magnetic-field applied in the $[111]$ direction, which had already been 
experimentally observed. 

In this publication, we study the evolution of the monopole density 
after a thermal quench. The description of such non-equilibrium dynamics is a 
worthwhile enterprise in itself, as thus far there have been no instances of 
three-dimensional magnets with pointlike elementary excitations, and hence 
little motivation for their study. However, there has been work on 
the quench dynamics of Coulomb liquids~\cite{Toussaint1983,Ginzburg1997},
which provides the starting point for our analysis. 

Our central result consists of the demonstration that the time-dependence of 
the monopole density after a quench provides a distinct signature of not only 
their pointlike nature but also of their magnetic Coulomb interaction.
For the nearest-neighbour model, we show that 
mean-field theory applies. We analytically account for the simulated 
time dependence of the density without free parameters. In dipolar spin ice, 
monopole bound states appear which can only be annihilated over an energy 
barrier. This leads to a dynamical arrest at low temperatures. 
This is again borne out by Monte Carlo simulations, where the fundamental 
dynamical move consists of a single spin flip as appropriate for the large 
Ising spins in spin ice~\cite{Jaubert2009}.

The outline of the paper is as follows. We set the stage by briefly 
summarising the annihilation-diffusion physics in Coulomb liquids. 
We address the new features present in nearest-neighbor spin ice, before 
presenting our results on the dipolar system. 
We close with remarks on equilibration in spin ice, and how the freezing of 
bound pairs could be used as an experimental technique to achieve measurable 
monopole densities at very low temperatures. 
%
%

\unue{Diffusion-annihilation in Coulomb liquids}
Consider a density $n_\pm (r)$ of positive and negative monopoles. 
As oppositely charged pairs can annihilate, their density obeys 
\beq
\frac{dn_+(r)}{dt} = \frac{dn_-(r)}{dt} = -\K n_+(r) n_-(r)
, 
\label{eq: dndt} 
\eeq
where $\K$ is an appropriate rate constant. In addition, the monopoles move 
deterministically in response to their mutual forces, and they are subject 
to diffusion in the presence of density inhomogeneities. 

Neglecting density fluctuations, 
one obtains the mean-field solution 
\bea
\rho(t) \equiv \frac{n_+ + n_-}{2} = \frac{\rho_0}{1 + \K \rho_0 t} 
\label{eq: mfres}
\eea 
for a quench to $T = 0$, where $\rho_0 \sim a_0^{-3}$ is the initial 
density: the characteristic timescale for the decay is 
$\tau_\K \sim a_0^3/\K$. 
A dynamical bottleneck can arise if there are spatial fluctuations in the 
relative density of positive and negative monopoles 
$\sigma(r) = [n_+(r)  - n_-(r)]/2$ 
(which is unaffected by the symmetric annihilation process) that are not 
smoothed fast enough by the motion of the monopoles. The relevant timescale 
for a particle to move a distance $a_0$ is given by 
$\tau_Q \sim a_0 / (\mu E) \sim a^3_0/(\mu q)$, where $\mu$ is the monopole 
mobility and $E \sim q/a^2_0$ is the typical strength of the Coulomb field. 
%
%

\unue{Nearest-neighbor spin ice} 
This system presents a number of special features with respect to ordinary 
Coulomb liquids, which are intricately linked to the existence of the monopoles 
against a backdrop of spin configurations in spin ice. The first is a
constraint on the possible values of $\sigma$ which follows from the fact that
the charge density encodes the change in the magnetisation of the sample. The
boundedness of the magnetisation implies that a cube of volume $L^3$ can at most
accommodate a net charge $\sigma \sim L^2$. Thus the long-wavelength Fourier
components are suppressed as $\tilde{\sigma} (q) \sim q^2$. 

\begin{figure}[ht]
\begin{center}
\includegraphics[width=0.49\columnwidth]{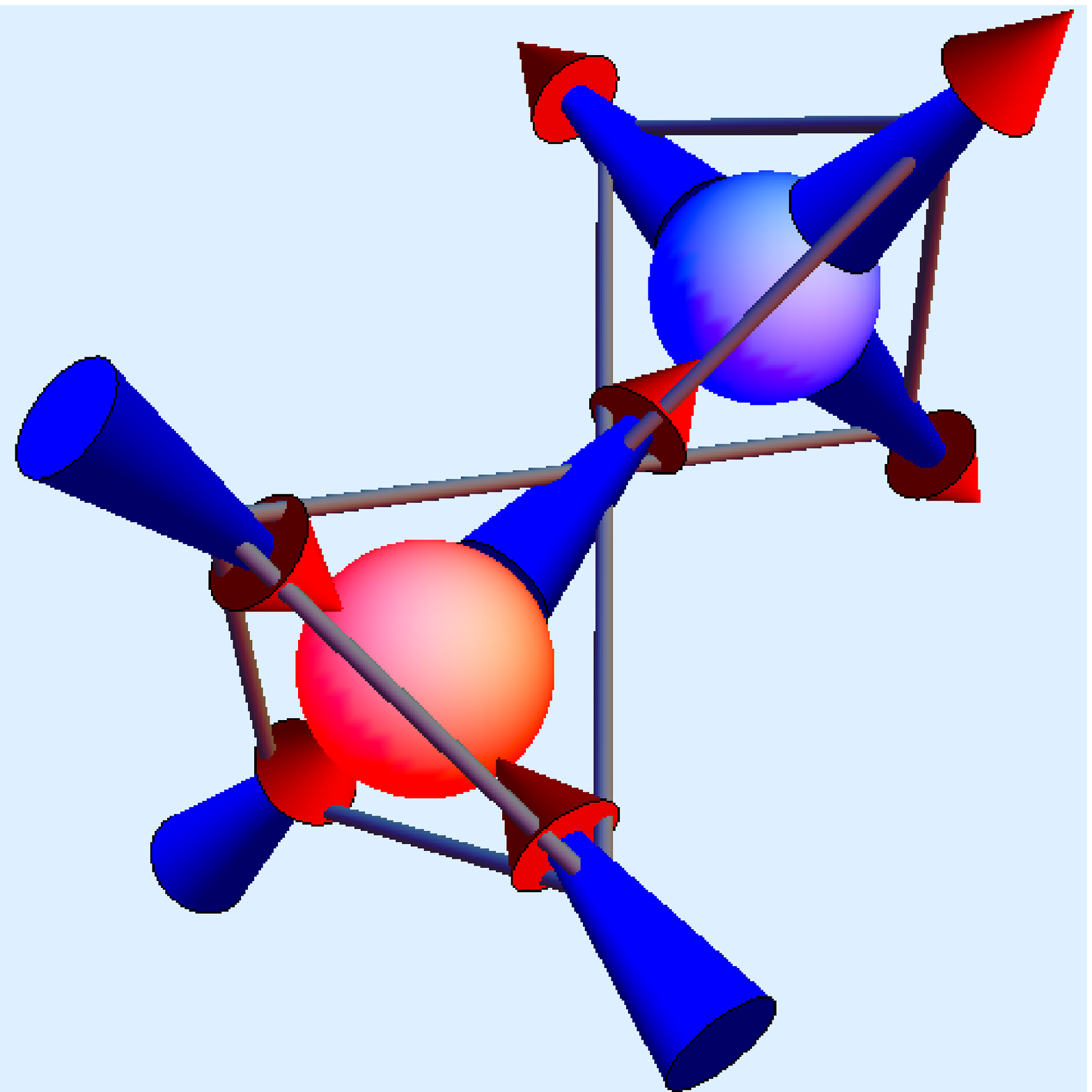}
\includegraphics[width=0.49\columnwidth]{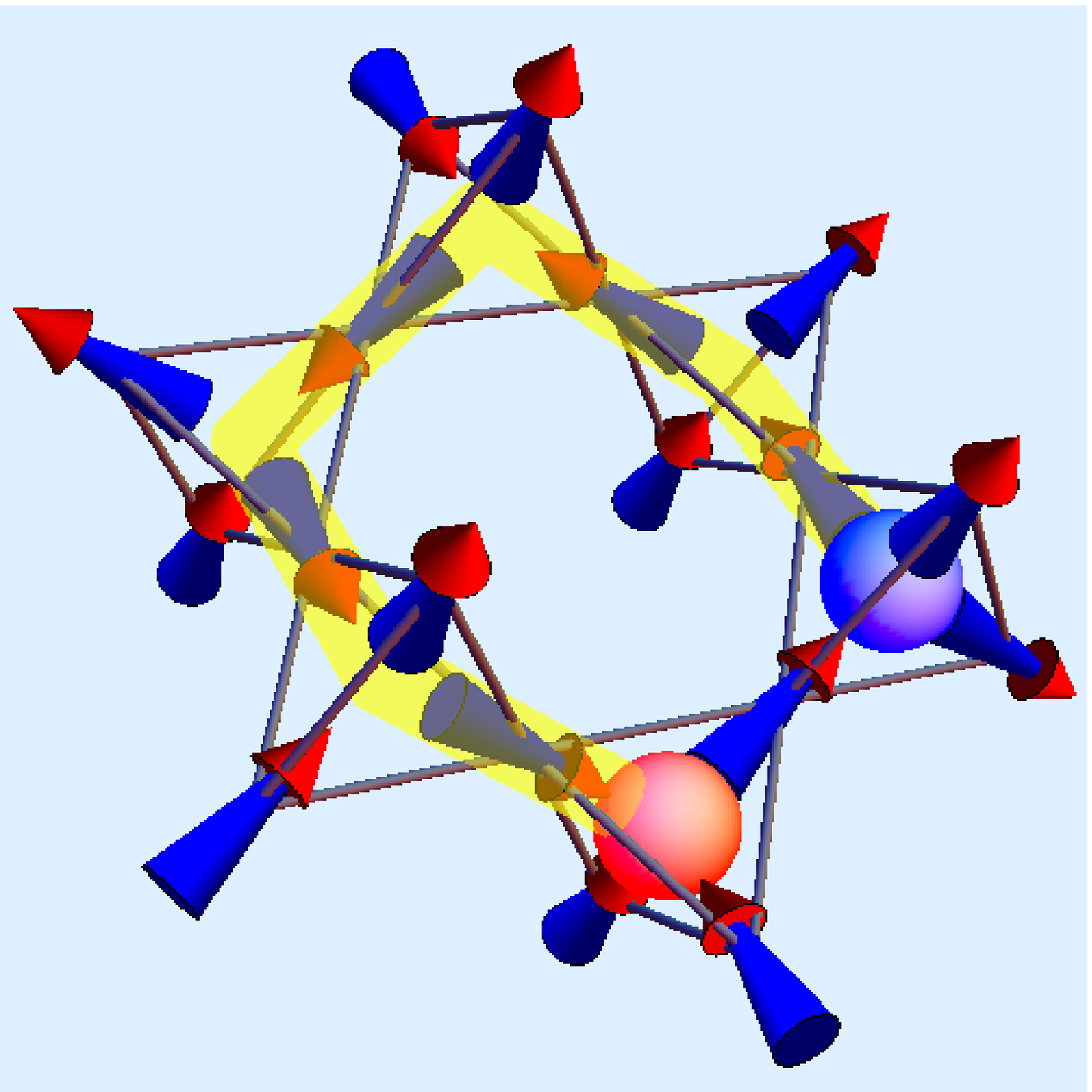}
\end{center}
\caption{
\label{fig: frozen pair} 
(Color Online) -- 
Example of a \emph{non-contractible} monopole-antimonopole pair (left panel). 
The shortest path that can lead to their annihilation is a hexagonal 
loop, provided the spins along the path are oriented appropriately 
(right panel). 
One can see explicitly that the two monopoles must separate before they are 
allowed to annihilate, resulting in a Coulomb energy barrier for the 
process. 
}
\end{figure}

The other crucial feature is that the interaction between the defects 
is of a purely entropic nature, due to the weighting of the monopole states by 
the number of spin configurations they are compatible with. 
This interaction has a Coulombic form
\beq
V_s (r) = k_B T \frac{Q^2_s}{r/a_d} 
, 
\label{eq: sigsT} 
\eeq
where $Q^2_s \simeq 0.35 \pm 0.01$ can e.g. be obtained from the
probability distribution of the separation of a lone pair of monopoles in
equilibrium Monte Carlo simulations~\cite{Krauth2003,Castelnovo2009}. 
Whereas the strength of this interaction vanishes as $T \to 0$, the mobility 
$\mu \simeq (Q_m a^2_d)/(6 \tau k_B T)$ arising from the single spin-flip 
Metropolis dynamics diverges, resulting in a regular $T \to 0$ limit of 
$\tau_Q$. 
Here $\tau$ is the basic unit of time, e.g., the inverse of the flip rate of 
an isolated spin in Monte Carlo simulations. 
In spin ice materials, AC susceptibility measurements seem to indicate that 
$\tau \sim 1$~ms~\cite{Snyder2004}.
A simple estimate yields $\K/a_d^3 \simeq 2g / \tau$, where 
$g \in \left[ \frac{3}{4}, \frac{9}{10} \right]$: 
the probability of finding two oppositely charged monopoles on neighbouring 
tetrahedra is proportional to $\rho^2$, and they annihilate in the next step 
(after time $\tau$) if 
flipping the intermediate spin restores the ice rules in both tetrahedra; 
this probability, which depends on the spin correlations, is estimated by $g$ 
($1-g$ being the probability that the two monopoles do not annihilate upon 
flipping the intermediate spin, as illustrated in the left panel of 
Fig.~\ref{fig: frozen pair}). 

\begin{figure}[ht]
\begin{center}
\includegraphics[width=0.98\columnwidth,height=0.6\columnwidth]
                {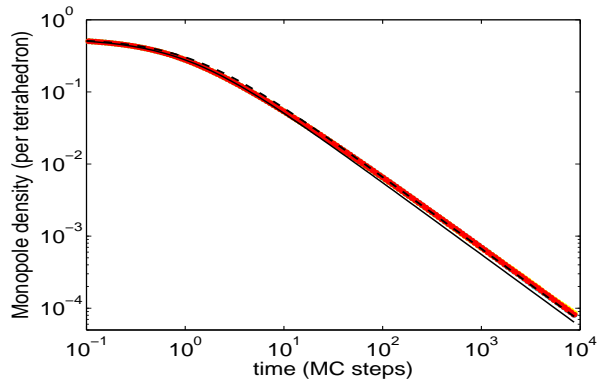}
\end{center}
\caption{
\label{fig: quench T=0 nn} 
(Color Online) -- 
Monopole density evolution in nearest-neighbour spin ice ($J=1$~K), after a 
temperature quench from $T=10$~K down to $T=0$~K. 
We simulates systems of size $L=32,\,64,\,128$, and finite size effects are 
absent at these time scales (estimated error bars are smaller than the 
symbol size). 
The analytical mean-field result Eq.~(\ref{eq: mfres}) 
is shown for $g = 3/4$ (dashed black line) and $g=9/10$ 
(solid black line). 
}
\end{figure}

Our numerical simulations of thermal quenches in nearest-neighbour spin ice
down to zero temperature, are displayed in Fig.~\ref{fig: quench T=0 nn}.
The mean field solution shows {quantitative} agreement with the 
numerics \emph{without any fitting parameters}: the fact that $\tilde{\sigma}
(q) \sim q^2$, together with the entropic Coulomb interaction, effectively 
suppress fluctuations in the charge density. 
%
%
%
%

\unue{Dipolar spin ice} 
The presence of a magnetic Coulomb interaction in spin ice leads to 
further features outside the conventional picture of Coulomb liquids. 
First of all, a diverging mobility is no longer compensated by a vanishing 
potential energy, and $\tau_Q \to 0$ in the zero temperature limit. 
This indicates that the motion of the monopoles does not follow linear 
response but rather the monopoles move along the local field direction at the 
maximum speed permitted by microscopic constraints, namely 
one step in time $\tau$. 
We thus expect monopoles to find each other very efficiently, and therefore a 
decay of $\rho$ which is at least as fast as in the nearest-neighbour case. 

This is, however, not what happens. 

The interplay between long-range
interactions and constraints imposed by the underlying spin degrees of freedom
leads to the formation of \emph{non-contractible} monopole pairs, and the 
system exhibits a dynamical arrest. Indeed, not all nearest-neighbour
monopole-antimonopole pairs can be annihilated by flipping the shared spin 
(see Fig.~\ref{fig: frozen pair}, left panel). 
Annihilation can then take place only if the two monopoles separate and meet 
again elsewhere in the lattice. However, due to their magnetic Coulomb
interaction, there is an energy barrier for such process, leading to an
activated Arrhenius behaviour in the monopole density relaxation. 

The smallest possible energy barrier determines the long time behaviour in 
the system. This is given by an elementary move where the monopoles of a bound 
pair annihilate around one of the adjacent hexagonal loops in the lattice 
(see Fig.~\ref{fig: frozen pair}, right panel). 
Two of the five spin flips involved 
in such process increase the distance between oppositely charged monopoles. 
A rough estimate for the concomitant energy gaps (see 
Ref.~\onlinecite{Castelnovo2008}) is given by the Coulomb interaction 
between the magnetic charges. From the nearest neighbour value 
$E_{\rm nn} = \mu_0 Q^2_m / 4 \pi a_d = 3.06$~K, we obtain the barrier to hop 
to second neighbour distance $a_{2n} = \sqrt{8/3}\,a_d$, 
$\Delta_1 = E_{\rm nn}(1-a_d/a_{2n}) = 1.19$~K, 
and the barrier to hop from second to third neighbour distance 
$a_{3n} = \sqrt{11/3}\,a_d$, 
$\Delta_2 = E_{\rm nn}(a_d/a_{2n}-a_{2n}/a_{3n}) = 0.28$~K, 
leading to an overall energy barrier of 
$\Delta=\Delta_1+\Delta_2=1.47$~K. 
In practice, the energy cost of a spin flip varies 
due to the effectively random fields set up by nearby bound pairs, leading 
to a broadened distribution of the $\Delta_i$. 

We ran extensive numerical Monte Carlo (MC) simulations treating the 
long range dipolar interaction via the Ewald summation 
technique,~\cite{deLeeuw1980} and using the Waiting Time Method 
(WTM)~\cite{WTM_refs} with single spin flip updates 
to access the long time regime~\cite{footnote MC order}. 
We prepare the system at equilibrium at the initial temperature of $10$~K; 
we then set the temperature to its quench value at time $t=0$, and we 
start the measurements. 

\begin{figure}[ht]
\begin{center}
\includegraphics[width=1.0\columnwidth]
                {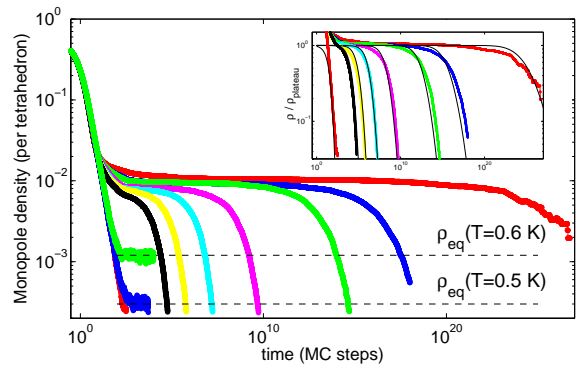}
\end{center}
\caption{
\label{fig: dipolar decay} 
(Color Online) -- 
Numerical simulations of thermal quenches in dipolar spin ice 
(system size $L=8$, i.e., $8192$ spins, and simulation parameters for 
{\DTO} as in Ref.~\onlinecite{denHertog2000}). 
The curves show the total density of defects $\rho$ per tetrahedron 
as a function of Monte Carlo time in units of Monte Carlo 
steps (1 attempt per spin), 
for quenches from $T=10$~K, down to 
$T=0.025$~K (red), 
$T=0.04$~K (blue), 
$T=0.05$~K (green), 
$T=0.075$~K (magenta), 
$T=0.1$~K (cyan), 
$T=0.125$~K (yellow), 
$T=0.15$~K (black), 
$T=0.4$~K (red), 
$T=0.5$~K (blue), 
and $T=0.6$~K (green) 
-- appearing in order from right to left. 
Inset: long time behaviour of $\rho$ normalised by its plateau 
value $\rho_{\rm plateau}$, compared to the phenomenological model discussed in 
the text (thin black lines). 
} 
\end{figure}

The defect density either reaches its equilibrium value very quickly, 
(for $T \gtrsim 0.4$~K), or a significant deviation from power law decay 
appears ($T \lesssim 0.4$~K) due to the activated behaviour induced by the 
non-contractible bound pairs, as illustrated in Fig.~\ref{fig: dipolar decay}. 

A (temperature independent) Gaussian distribution of energy barriers $\Delta$ 
peaked around $1.47$~K, with a variance $0.01$~K$^2$, leads to a probability 
distribution $\P(\Theta)$ of single hexagon decay times $\Theta$, and hence 
a (normalised) defect density $\rho(t) = 1 - \int^t_0 \P(\Theta) \:d \Theta$. 
The resulting curves $\rho(t)$ are compared with the numerical ones in the 
inset of Fig.~\ref{fig: dipolar decay}. Notice the good agreement over 
more than $20$ orders of magnitude in $t$, for the different values of the
quench temperature. 
Clearly, this phenomenological model captures the fundamental physics 
underlying the dynamical arrest in thermal quenches. 

\begin{figure}[ht]
\begin{center}
\includegraphics[width=0.98\columnwidth,height=0.6\columnwidth]
                {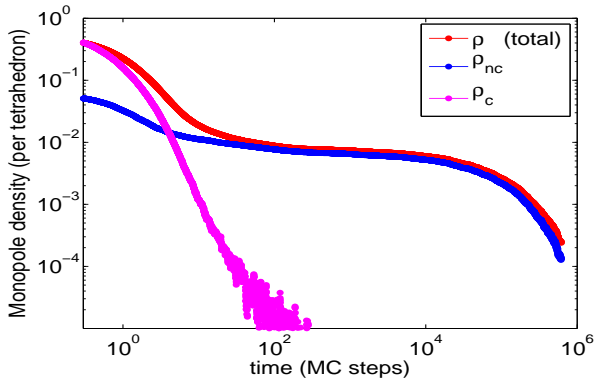}
\end{center}
\caption{
\label{fig: dipolar decay Tq0125} 
(Color Online) -- 
Numerical simulation of a thermal quench down to $T=0.125$~K (system size
$L=8$).  
The red curve shows the total density of defects per tetrahedron 
$\rho$, while the blue and the rapidly decaying magenta curves correspond to 
the density of defects forming non-contractible pairs $\rho_{\rm nc}$ and 
contractible pairs $\rho_{\rm c}$, respectively. 
} 
\end{figure}

To further confirm this scenario, we explicitly determined the density of 
monopoles forming non-contractible pairs,  
as well as the density of contractible defect pairs (i.e., pairs where 
flipping the intermediate spin lowers the number of defects in the system). 
The result is illustrated in Fig.~\ref{fig: dipolar decay Tq0125}, 
for a given quench temperature. 
One can see that the initial decay ends when there are 
essentially no contractible pairs left in the system (magenta curve falling 
below $1/N_t$, where $N_t = 8L^3$ is the total number of tetrahedra in the 
lattice). 
{}From thereon, the total defect density is essentially given by monopoles 
forming non-contractible pairs. 

\begin{figure}[ht]
\begin{center}
%
\includegraphics[width=0.8\columnwidth]{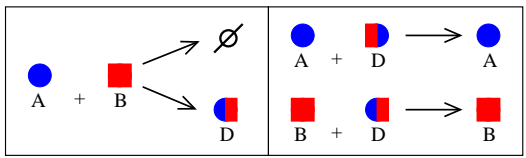}
\end{center}
\caption{
\label{fig: frozen pair mean field} 
(Color Online) -- 
Schematic illustration of the reaction processes in the mean field 
model used to describe the very low temperature limit of 
thermal quenches in spin ice. 
} 
\end{figure}

The defect density decay approaching the plateau is captured by 
a diffusion process where oppositely charged particles ($A$, $B$) can 
either annihilate ($\emptyset$) or fuse into a non-contractible pair ($D$) 
(Fig.~\ref{fig: frozen pair mean field}, left panel). 
For quenches to very low temperatures, the non-contractible pairs can be
approximated  
as frozen unless another single particle annihilates one member of the pair,
thus freeing the other one 
(Fig.~\ref{fig: frozen pair mean field}, right panel). 
In a simple  mean field model, one finds a surviving population of 
non-contractible pairs $D$, 
provided the single particle density decays faster than $1/t$, as is the case 
in our simulations. Indeed, the resulting time-dependence of the total 
and non-contractible particle densities is in good qualitative agreement with 
the numerical results illustrated in 
Fig.~\ref{fig: dipolar decay Tq0125}~\cite{Castelnovo2009}. 
On the longest timescales, annihilation of non-contractible pairs $D \to 0$ 
around hexagonal loops terminates the plateau. 
%
%

\unue{Equilibration timescales, and experiment} 
The dominant dynamics in spin ice at low temperatures consists of hopping 
monopoles. These correspond to single spin flips which do not incur a 
cost for violating the ice rules. As the temperature is lowered to zero, the 
monopole density vanishes and spin ice freezes completely. 
At finite temperatures the low density of monopoles leads to an exponentially 
large timescale (in fact, possibly super-exponentially 
large~\cite{Snyder2004,Jaubert2009}) which grows faster than the timescale 
governing the monopole density. 
Upon cooling, there is a ``fast'' process responsible for the thermalisation 
of the energy (i.e., the monopole density), and  a much slower 
process that equilibrates the spin correlations. We believe that this 
mechanism explains why certain quantities such as the energy seem to 
equilibrate at temperatures where the magnetisation has long fallen out of 
equilibrium. It would be on the long time scales of the slower process 
that any magnetic order would be established~\cite{Melko2001}. 

In addition, we note that non-contractible bound monopoles are practically 
absent upon heating from an equilibrium state. There is thus an asymmetric 
approach to equilibrium in the monopole density at a given temperature, 
depending on whether we use heating or cooling thermal quenches. 

The most elegant way to measure monopole densities would be via zero-field 
NMR on the oxygen nuclei at the centre of the tetrahedra, which experience 
different field strengths and corresponding fluctuation rates when the 
tetrahedra host a monopole \cite{takiHFM}.
The aim is to have a sufficiently large density 
of monopoles to yield a measurable signal, while preventing the monopoles 
from moving around too quickly, thus spoiling the measurement. 
In thermal equilibrium, temperatures low enough for the 
second condition to be satisfied result in exceedingly small monopole
densities.  
Our results suggest that a temperature quench on time scales sufficiently 
shorter than the time to develop the dynamical arrest plateau in 
Fig.~\ref{fig: dipolar decay} ($\sim 10$-$100$~ms) could be used to induce a 
monopole-rich state ($\rho \gtrsim 10^{-2}$ per tetrahedron), 
where motion at short times is obstructed by 
the formation of non-contractible pairs. 

Finally, the method of choice for imaging spin correlations is neutron 
scattering~\cite{Morris2009,Fennell2009,Kadowaki2009}. 
As the non-contractible monopole pairs remain bound on long timescales, the 
concomitant short-range correlations should be visible in the neutron 
scattering cross section. 
%
%

\unue{Acknowledgements}
We are grateful to Steve Bramwell and Martin Zapotocky for helpful discussions, 
and to Paul Krapivski for suggesting the mean field model 
with completely frozen pairs. 
This work was supported in part by EPSRC Grant No. GR/R83712/01 (CC). 
%
%

\end{document}